\documentclass[prd,twocolumn,superscriptaddress]{revtex4}
\usepackage{amsmath,amssymb,epsfig,bm,mathrsfs,feynmp,color}
\usepackage{slashed}
\newcommand{\be}{\begin{equation}}
\newcommand{\ee}{\end{equation}}
\newcommand{\bea}{\begin{eqnarray}}
\newcommand{\eea}{\end{eqnarray}}

\def\XXint#1#2#3{{\setbox0=\hbox{$#1{#2#3}{\int}$}
     \vcenter{\hbox{$#2#3$}}\kern-.5\wd0}}

\definecolor{red}{rgb}{0.8,0,0}
\definecolor{orange}{rgb}{0.8,0.2,0.0}
\definecolor{blue}{rgb}{0.3,0.0,0.8}

\begin{document}


\title{Radial oscillations of neutral and charged hybrid stars}

\author{Alessandro Brillante} 
\affiliation{Frankfurt Institute for Advanced Studies (FIAS), Ruth-Moufang-Stra{\ss}e 1, 60438 Frankfurt am Main, Germany}
\affiliation{Institut f\"{u}r Theoretische Physik, Johann Wolfgang Goethe-Universit\"{a}t, Max-von-Laue-Stra{\ss}e 1, 60438 Frankfurt am Main, Germany }

\author{Igor N. Mishustin}
\affiliation{Frankfurt Institute for Advanced Studies (FIAS), Ruth-Moufang-Stra{\ss}e 1, 60438 Frankfurt am Main, Germany}
\affiliation{Kurchatov Institute, Russian Research Center, Akademika Kurchatova 1, 123182 Moscow, Russia}
\begin{abstract}
We construct stellar models of hadron stars and hybrid stars and calculate the frequencies of their lowest radial mode of vibration. Chandrasekhar's equation for radial oscillations is generalized for stars with internal electric fields and earlier versions of that generalization are simplified. For the hybrid stars a Gibbs construction is employed. It is found that the softening of the equation of state associated with the presence of deconfined quarks reduces the oscillation frequency. We show that a slight charge inbalance should lead to increased maximum mass, decreased central density and lower oscillation frequencies.
\end{abstract}

\maketitle

\section{Introduction}
The research on stellar oscillations within the framework of GR goes back to the early studies of Chandrasekhar from almost half a century ago \cite{chandra}. The field continues to attract great interest, because the investigation of stellar oscillations can reveal the inner structure of dense stars and shed light on the equation of state of superdense matter. In this paper we focus exclusively on radial oscillations, for which the assumption of spherical symmetry holds. While radial oscillations are not detectable by gravitational waves due to the vanishing quadrupole moment of a radially moving sphere, they are in principle observable by emission of electromagnetic radiation, whenever there is a charge separation at the surface of the star. First calculations for radial modes with a sample of different EoS were performed in \cite{harrison}\cite{chanmugam}\cite{glass}. A large survey of radial oscillations for several zero-temperature equations of state was done in \cite{kokkotas}. The radial modes of strange stars were studied in \cite{benvenuto}\cite{lugones} and references therein. In \cite{lugones} the effect of color-flavor locking in the quark phase of strange stars on the radial modes was investigated. They found, that the oscillation periods for high mass strange stars are shifted to higher values, as the gap parameter is increased. Nonlinear effects in radial oscillations were discussed in \cite{gabler}, where it was found that they may stabilize stars beyond the "maximum mass limit", due to nonlinear coupling of modes.

This paper complements the existing studies on radial modes in several respects. First, we consider stellar models which give maximum masses close to the present observational limit of $2 M_{\odot}$ \cite{demorest}\cite{antoniadis}. Second, we calculate the radial modes for stars with a Gibbs construction and consider realistic equations of state for the quark and hadronic phases \cite{alford2}\cite{weissenborn}. Some of the previous studies on radial modes for hybrid stars \cite{flores2} used the Maxwell construction, where there is a discontinuity in the baryon density. As demonstrated in \cite{mishustin2} the models with Maxwell construction lead to higher maximum masses then their Gibbs counterparts. Generally, we find a reduction in the frequencies associated with the deconfinement phase transition.

In our analysis we also consider compact stars with net electric charge. Of course, we assume that the net charge is rather small, so that the star is not destroyed by the Coulomb force. As shown in \cite{alcock}\cite{malheiro}\cite{ebel} even globally neutral stars can develop strong surface electric fields. For this purpose we rederive the charge-generalization of the pulsation equation. This equation has already appeared in different notations in the literature. In the present paper we provide a clear derivation and obtain a simpler expression than those given in \cite{glazer1}\cite{rothman}.
Below we apply the derived oscillation equation and present first numerical results obtained with the correct equation. In contrast to earlier studies we investigate charged stars with a realistic EoS. Our results show that the presence of a slight charge inbalance may help to reconcile soft equations of state with observations of massive pulsars. Furthermore, an enhancement of the oscillation period at given central baryon density is observed.
\section{Equation of state for hybrid stars}
\noindent Presently it is not possible to describe nuclear matter and deconfinied quark matter in an unified approach. We are forced to calculate each phase separately and match them by a phase transition. Here we will employ the Gibbs construction, which corresponds to a small surface tension of nuclear matter. Within the Gibbs approach there is a coexistence region, where both phases coexist with varying proportions. This is in contrast to the Maxwell construction, where two pure phases meet at one boundary. In contrast to the Maxwell approach, we do not encounter a discontinuity in the density. This is crucial for the applicapility of the perturbative approach, which requires the smallness of all perturbed quantities, including the fluid pressure and density. The Gibbs condition is more involved, because there are 2 (ore even more) independent chemical potentials, each corresponding to one globally conserved charge. In this case the Gibbs condition reads
\begin{align}
P_1(\mu_B,\mu_Q)&=P_2(\mu_B,\mu_Q)\notag\\
\mu_{B1}&=\mu_{B2}\notag\\
\mu_{Q1}&=\mu_{Q2}\notag\;,
\end{align} where $\mu_B$ and $\mu_Q$ are the independent chemical potentials.
For the hadronic phase the relativistic mean field model is used, which provides an effective and relativistically invariant description for dense multi-particle systems. The parameters are fitted to the known properties of nuclear matter at saturation. We use the TM1 parameter set, which leads to a maximum mass of about $2.2M_{\odot}$\cite{sugahara}. The quark phase is described by a MIT bag model with the free energy \cite{alford2}
\begin{align}
\Omega&=\sum_{i=u,d,s,e} \Omega_i+\frac{3\mu^4}{4\pi^2}\left(1-a_4\right)+B\;,
\end{align}
where the first term represents the thermodynamic potentials of non-interacting particles, $a_4$ is the interaction strength and $B$ is the bag constant.

\section{Charge density distribution}
Below we consider both neutral and charged stars. In the latter case there is less certainty about the distribution of charge density $\rho_{ch}$ than about the EoS. While for constant density stars $\rho_{ch}$ might increase monotonically toward the surface, this distribution is inconsistent with the employed EoS described above. We assume that $\rho_{ch}$ is proportional to the baryon number density. The total charge is assumed to be small in the sense that the corresponding electrostatic energy is smaller than the gravitational energy, i.e.
\begin{align}
\frac{{N_C}^2 e^2}{R}<\frac{G {N_B}^2 {m_p}^2}{R}, \;\; \frac{N_C}{N_B}<\sqrt{\frac{G{m_p}^2}{e^2}}<10^{-18}\;,
\end{align}
where $N_B$ and $N_C$ denote the total baryon number and charge number, respectively. Below we use the parameter $x=10^{19} N_C/N_B$, measuring the charge of the star.

\section{Oscillation equation}
\noindent In this section we consider radial oscillations of compact stars. We briefly sketch the derivation of the pulsation equation following Chandrasekhar's work \cite{chandra}. Additionally we allow for a non-vanishing net charge.

The line element of the Schwarzschild-like metric is given by
\begin{equation}
ds^2=-e^{2\Phi}dt^2+e^{2\Lambda}dr^2+r^2d{\theta}^2+r^2sin^2\theta d{\phi}^2\label{C5}
\end{equation}
where the metric potentials $\Phi$ and $\Lambda$ are only functions of $x^0=t$ and $x^1=r$. Then the Einstein tensor is given by
\begin{align}
G_0^{\;\;0}&=-e^{-2\Lambda}\left[2r^{-1}\Lambda'-\left(1-e^{2\Lambda}\right)r^{-2}\right]\label{C6}\\
G_1^{\;\;1}&=e^{-2\Lambda}\left[2r^{-1}\Phi'+r^{-2}\right]-r^{-2}\label{C7}\\
G_2^{\;\;2}&=e^{-2\Lambda}\left[\Phi''-\Phi'\Lambda'+{\Phi'}^2+r^{-1}\left(\Phi'-\Lambda'\right)\right]\notag\\&+e^{-2\Phi}\left[\dot{\Phi}\dot{\Lambda}-\ddot{\Lambda}-{\dot{\Lambda}}^2\right]\label{C8}\\
G_0^{\;\;1}&=2r^{-1}e^{-2\Lambda}\dot{\Lambda}\label{C9}\;,
\end{align}
where $a'$ and $\dot{a}$ denote $\partial_r a$ and ${\partial_t a}$, respectively. The stress-energy tensor is given by
\begin{align}
T_{\mu}^{\;\;\nu}&=\left(\rho+P\right)u_{\mu}u^{\nu}+Pg_{\mu}^{\;\;\nu}\notag\\&+\frac{1}{4\pi}\left[F_{\mu\alpha}F^{\alpha\nu}-\frac{1}{4}g_{\mu}^{\;\;\nu}F^{\beta\gamma}F_{\beta\gamma}\right]\label{C10}\;,
\end{align}
where $\rho$, $P$ and $F$ denote the fluid energy density, fluid pressure and the Maxwell tensor, respectively.
Due to spherical symmetry, $F_{10}=-F_{01}=E_r$ is the only nonvanishing component of $F$. For the Schwarzschild metric the enclosed electric charge is defined as $Q\equiv4\pi\int_0^r {r'}^2e^{\Phi+\Lambda}j^0dr'$. Maxwell's equation $\partial_{\mu}\left[\sqrt{-g}F^{\nu\mu}\right]=4\pi\sqrt{-g}j^{\nu}$ relates the enclosed charge to the electric field $Q=r^2e^{-\Phi-\Lambda}E$. Therefore the differential equations for the charge are given by
\begin{align}
\dot{Q}&=-4\pi r^2\rho_{ch}e^{\Phi+\Lambda}u^1\label{Qt1}\\
Q'&=4\pi r^2\rho_{ch}e^{\Phi+\Lambda}u^0\label{Q1}\;,
\end{align}
where \eqref{Qt1} follows from Maxwell's equation and \eqref{Q1} is true by definition. The nontrivial components of the covariant divergence of the stress-energy tensor are
\begin{align}
T_{0\;\;;\nu}^{\;\nu}&=\partial_0 T_0^{\;0}+\partial_1 T_0^{\;1}+\dot{\Lambda}\left(T_0^{\;0}-T_1^{\;1}\right)\notag\\&+\left(\Phi'+\Lambda'+2r^{-1}\right)T_0^{\;1}\label{C15}\\T_{1\;\;;\nu}^{\;\nu}&=\partial_0 T_1^{\;0}+\partial_1 T_1^{\;1}-\left(\dot{\Phi}+\dot{\Lambda}\right)e^{2\Lambda-2\Phi}T_0^{\;1}\notag\\&+\Phi'\left(T_1^{\;1}-T_0^{\;0}\right)+2r^{-1}\left(T_1^{\;1}-P-\frac{Q^2}{8\pi r^4}\right)\label{C16}\;.
\end{align}
In order to evaluate Einstein's equation at equilibrium, we impose stationarity and set the radial velocity to zero.
\begin{align}
{\Lambda_0}'&=4\pi r\rho_0 e^{2\Lambda_0}+\frac{1-e^{2\Lambda_0}}{2r}+\frac{{Q_0}^2e^{2\Lambda_0}}{2r^3}\label{C19}\\
{\Phi_0}'&=4\pi rP_0e^{2\Lambda_0}+\frac{e^{2\Lambda_0}-1}{2r}-\frac{{Q_0}^2e^{2\Lambda_0}}{2r^3}\label{C20}\\
{P_0}'&=-\left(\rho_0+P_0\right){\Phi_0}'+\frac{Q_0 {Q_0}'}{4\pi r^4}\label{C21}\\
{\Phi_0}''&={\Phi_0}'{\Lambda_0}'-{{\Phi_0}'}^2+r^{-1}\left({\Lambda_0}'-{\Phi_0}'\right)\notag\\&+8\pi e^{2\Lambda_0}P_0+\frac{e^{2\Lambda_0}{Q_0}^2}{r^4}\label{C57}
\end{align}
\eqref{C21} is the condition of hydro-electrostatic equilibrium derived by Bekenstein in \cite{bekenstein}. Neglecting nonlinear terms in the perturbations, the 4-velocity can be written as
\begin{equation}
u^0=e^{-\Phi}\;\;u^1=ve^{-\Phi_0}\;\;u_0=-e^{\Phi}\;\;u_1=ve^{2\Lambda_0-\Phi_0}\label{C25}\;,
\end{equation}
where $v$ is defined as
\begin{align}
v=\frac{dx^1}{dx^0}\label{C26}\;.
\end{align}
As pointed out in \cite{knutsen}, it is not justified to replace $\Phi$ by $\Phi_0$ in the expressions for $u^0$, $u_0$. If we subtract from \eqref{C16} the same equation evaluated at equilibrium and keep only terms linear in the perturbations, we obtain
\begin{align}
&e^{2\Lambda_0-2\Phi_0}\left(\rho_0+P_0\right)\dot{v}+{\delta P}'+\frac{Q_0{Q_0}'\xi'}{4\pi r^4}\notag\\&+\frac{Q_0{Q_0}''\xi}{4\pi r^4}+\frac{{{Q_0}'}^2\xi}{4\pi r^4}+{\Phi_0}'\left(\delta\rho+\delta P\right)\notag\\&+\left(\rho_0+P_0\right)\delta\Phi'=0\label{C33}\;.
\end{align}
$\xi=r-r_0$ is the absolute displacement of a fluid element relative to its equilibrium position. The time derivative of the displacement $\xi$ is equal to the radial velocity $v$.
\begin{align}
v&=\frac{d\xi}{dt}
\end{align}
Assuming that the perturbations obey an harmonic time dependence $e^{i \omega t}$ \eqref{C33} reduces to
\begin{align}
&e^{2\Lambda_0-2\Phi_0}\left(\rho_0+P_0\right)\omega^2\xi={\delta P}'+\frac{Q_0{Q_0}'\xi'}{4\pi r^4}\notag\\&+\frac{Q_0{Q_0}''\xi}{4\pi r^4}+\frac{{{Q_0}'}^2\xi}{4\pi r^4}+{\Phi_0}'\left(\delta\rho+\delta P\right)\notag\\&+\left(\rho_0+P_0\right)\delta\Phi'=0\label{C43}\;,
\end{align} where the variables for the perturbations now denote their time-independent amplitudes.
\eqref{C43} is the prototype of the pulsation equation. What remains to be done, is to insert in \eqref{C43} the expressions for $\delta\rho$, $\delta P$, $\delta P'$ and $\delta\Phi'$.
If we inspect the only offdiagonal component of Eintein's equation, keep only linear terms and integrate, we arrive at
\begin{align}
\delta\Lambda&=-\left({\Phi_0}'+{\Lambda_0}'\right)\xi\label{C36}\;.
\end{align}
The diagonal components of Einstein's equation can be subtracted from their equilibrium values. After linearization, one obtains
\begin{align}
\delta\rho&=-\xi{\rho_0}'-\left(\rho_0+P_0\right)\frac{e^{\Phi_0}}{r^2}\left(r^2 e^{-\Phi_0}\xi\right)'\label{C39}\\
\delta\Phi'&=4\pi re^{2\Lambda_0}\delta P+2{\Phi_0}'\delta\Lambda+r^{-1}\delta\Lambda\notag\\&-\frac{Q_0{\delta Q}e^{2\Lambda_0}}{r^3}\label{C41}\;,
\end{align}
where the charge-dependence of $\delta\rho$ in \eqref{C39} is only implicit. We assume a barotropic equation of state $P=P\left(\rho\right)$ and define the adiabatic index as
\begin{align}
\gamma&=\frac{\rho_0+P_0}{P_0}\frac{d P_0}{d \rho_0}\label{C54}
\end{align}
Therefore $\delta P$ is already constrained as
\begin{align}
\delta P&=\frac{d P_0}{d\rho_0}\delta\rho=-\xi {P_0}'-\frac{\gamma P_0 e^{\Phi_0}}{r^2}\left(r^2e^{-\Phi_0} \xi\right)'\label{C53}
\end{align}
If the equation of state is not barotropic and the pressure is a function of the energy density and baryon density, it is possible to define the adiabatic index in such a way \cite{chandra}, as to preserve the relationship between $\delta\rho$ and $\delta P$. Therefore the pulsation equation does not rely on this assumption. With the previous equations, \eqref{C43} reduces to
\begin{align}
&\omega^2 e^{2\Lambda_0-2\Phi_0}\left(\rho_0+P_0\right)\xi=-e^{-\Lambda_0-2\Phi_0}\bigg[e^{\Lambda_0+3\Phi_0}\frac{\gamma P_0}{r^2}\notag\\&\left(r^2e^{-\Phi_0}\xi\right)'\bigg]'-\left(\rho_0+P_0\right){\Phi_0'}^2\xi+4r^{-1}\xi{P_0}'\notag\\&+8\pi\left(\rho_0+P_0\right)\xi e^{2\Lambda_0}P_0+\left(\rho_0+P_0\right)r^{-4}\xi e^{2\Lambda_0}{Q_0}^2\label{C59}\;.
\end{align}
This pulsation equation is equivalent to equation (28) in \cite{glazer1} and equation (4.13) in \cite{rothman}, but in contradiction to the corresponding equation in \cite{felice}, which seems to be wrong. We prefer our version, because it is more compact and captures the difference from Chandrasekhar's equation in a single term.

\begin{figure}[!t]
\begin{center}
\includegraphics[height=8.4cm,width=8.4cm,angle=0]{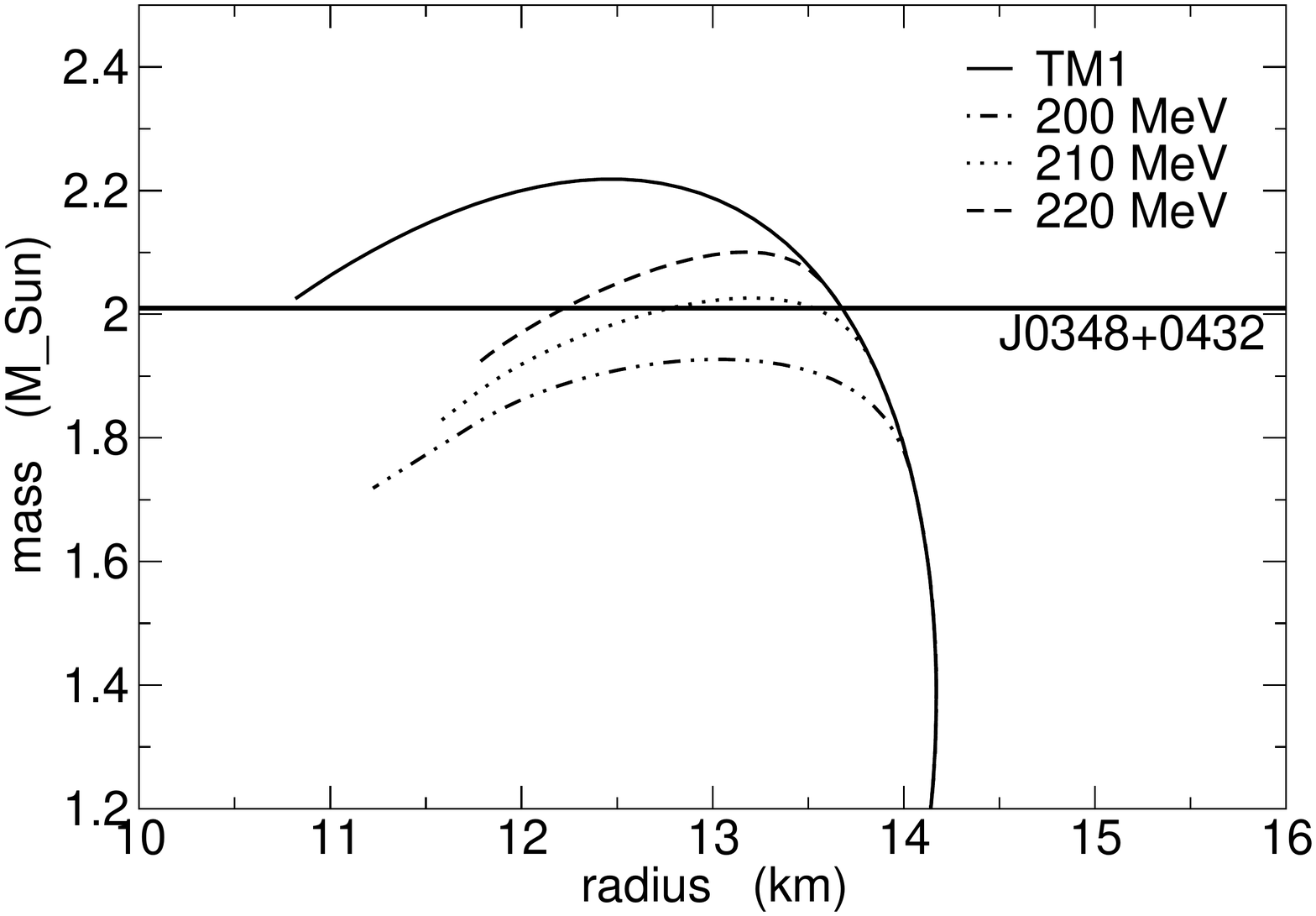}
\hskip 1.2cm
\includegraphics[height=8.4cm,width=8.4cm,angle=0]{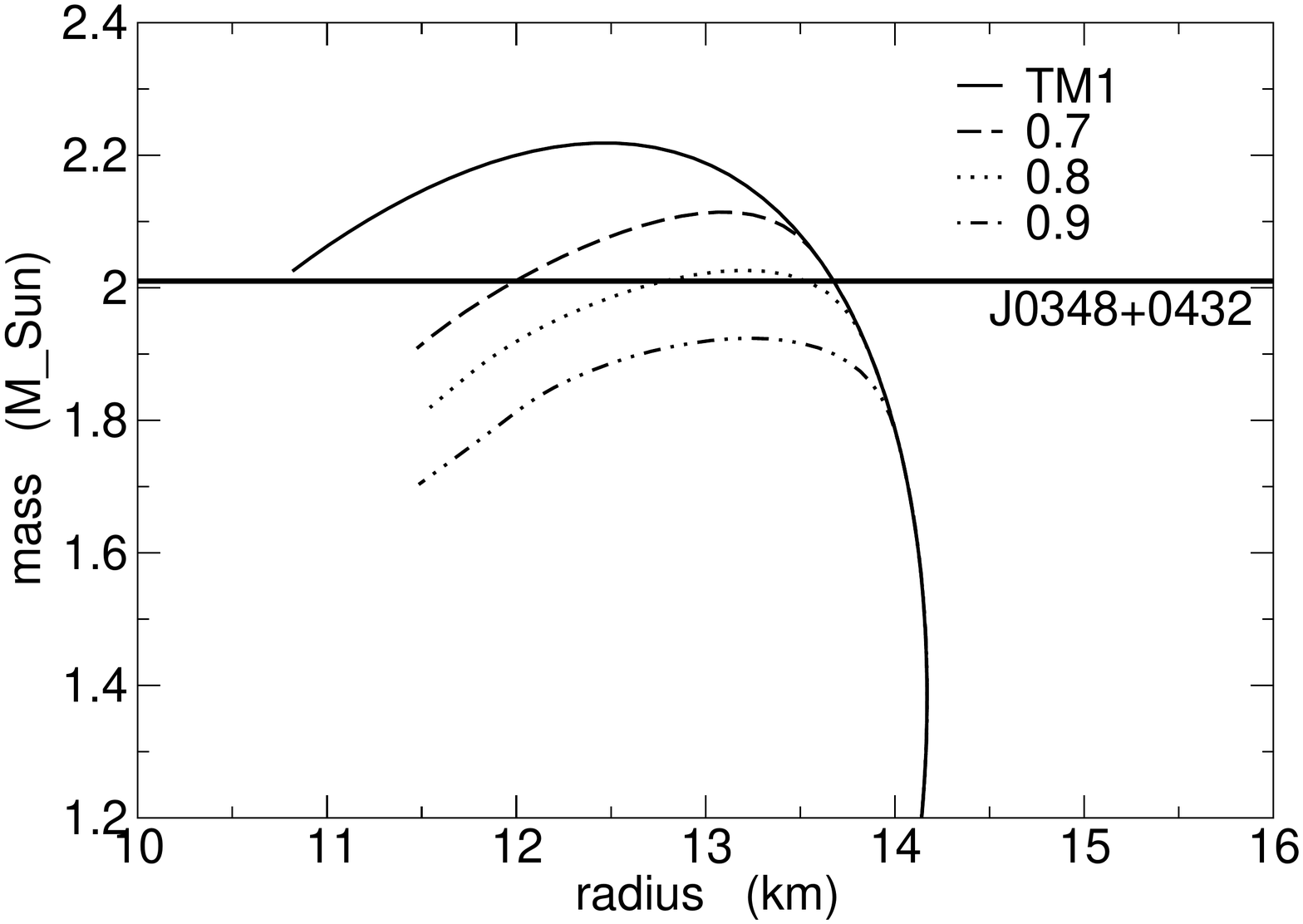}
\end{center}
\caption[]
{Mass-radius diagram of hybrid stars with a Gibbs construction. The lower limit for the maximum mass given by PSR J0348+0432 \cite{antoniadis} is indicated. In the upper pannel the parameters for the quark phase are $m_s=100\;MeV$, $a_4=0.8$ and $B^{\frac{1}{4}}$ as indicated in the plot. In the lower pannel the parameters are $m_s=100\;MeV$  $B^{\frac{1}{4}}=210\;MeV$ and $a_4$ as indicated in the plot.}
\label{fig:m-r_0}
\end{figure}
\begin{figure}[!t]
\begin{center}
\includegraphics[height=8.4cm,width=8.4cm,angle=0]{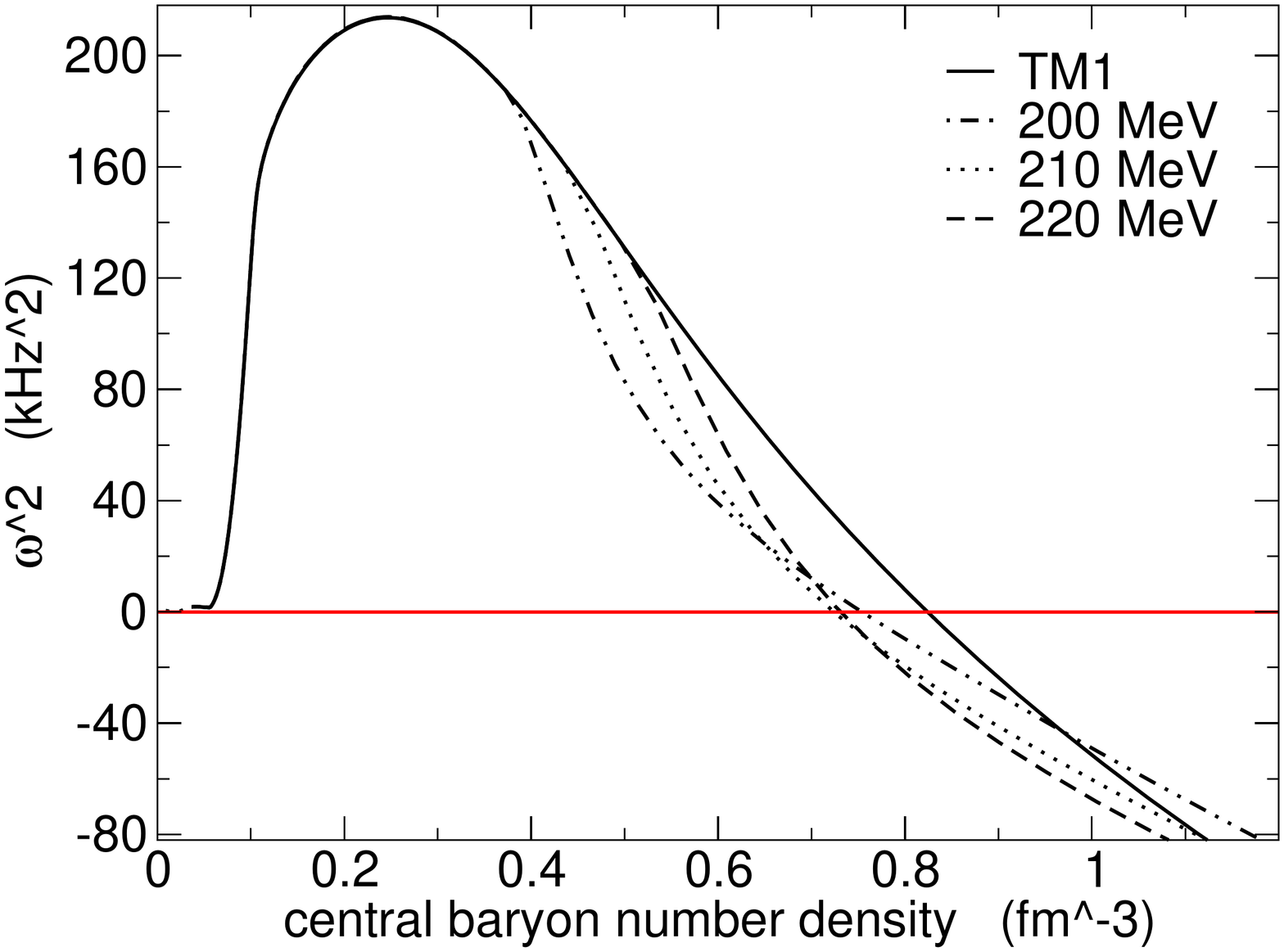}
\hskip 1.2cm
\includegraphics[height=8.4cm,width=8.4cm,angle=0]{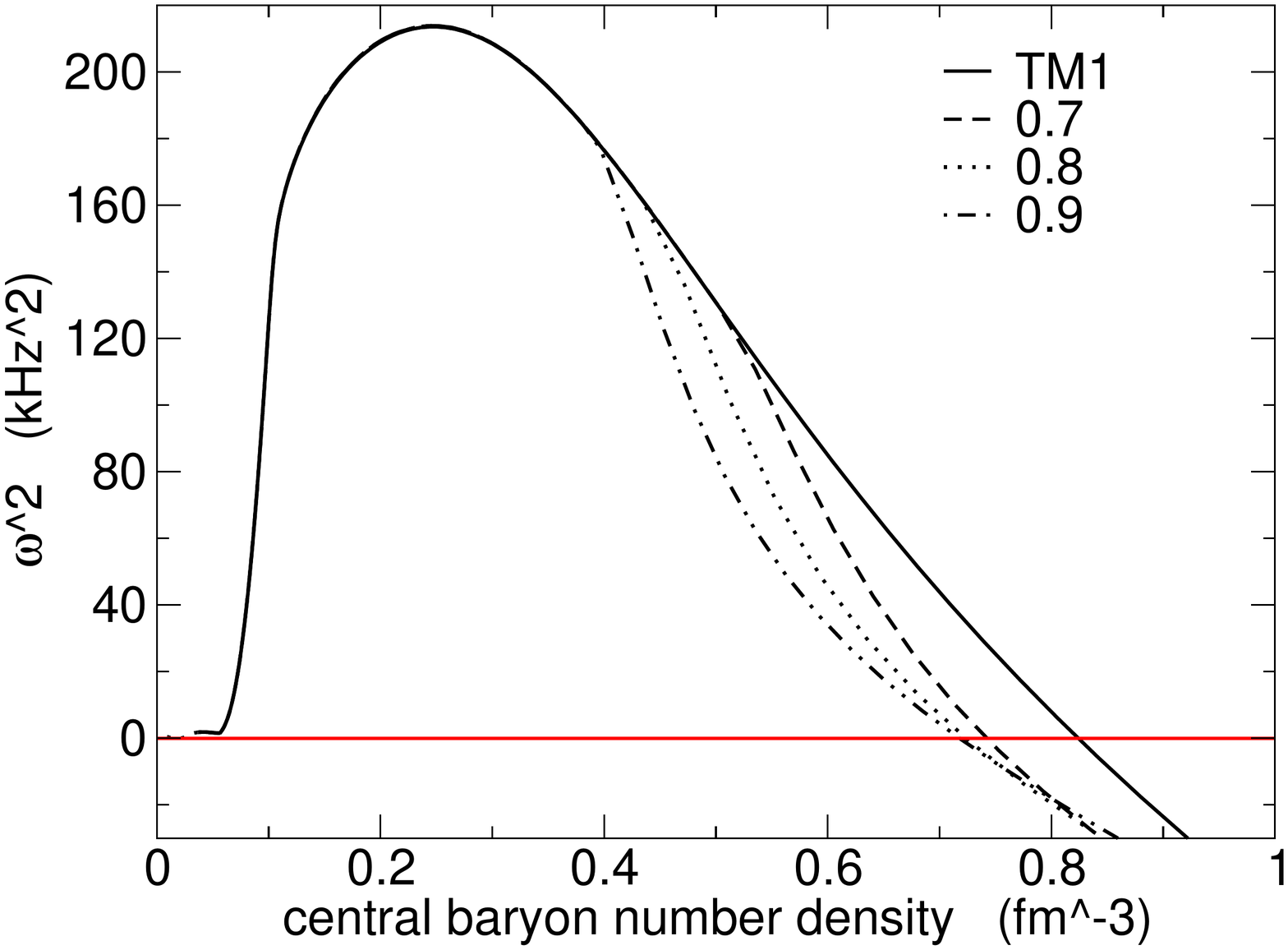}
\end{center}
\caption[]
{Square of the oscillation frequency of the fundamental mode as function of the central baryon number density. The parameter choices for both pannels are the same as in Figure \ref{fig:m-r_0}.}
\label{fig:osc_0}
\end{figure}

\begin{figure}[!t]
\begin{center}
\includegraphics[height=8.4cm,width=8.4cm]{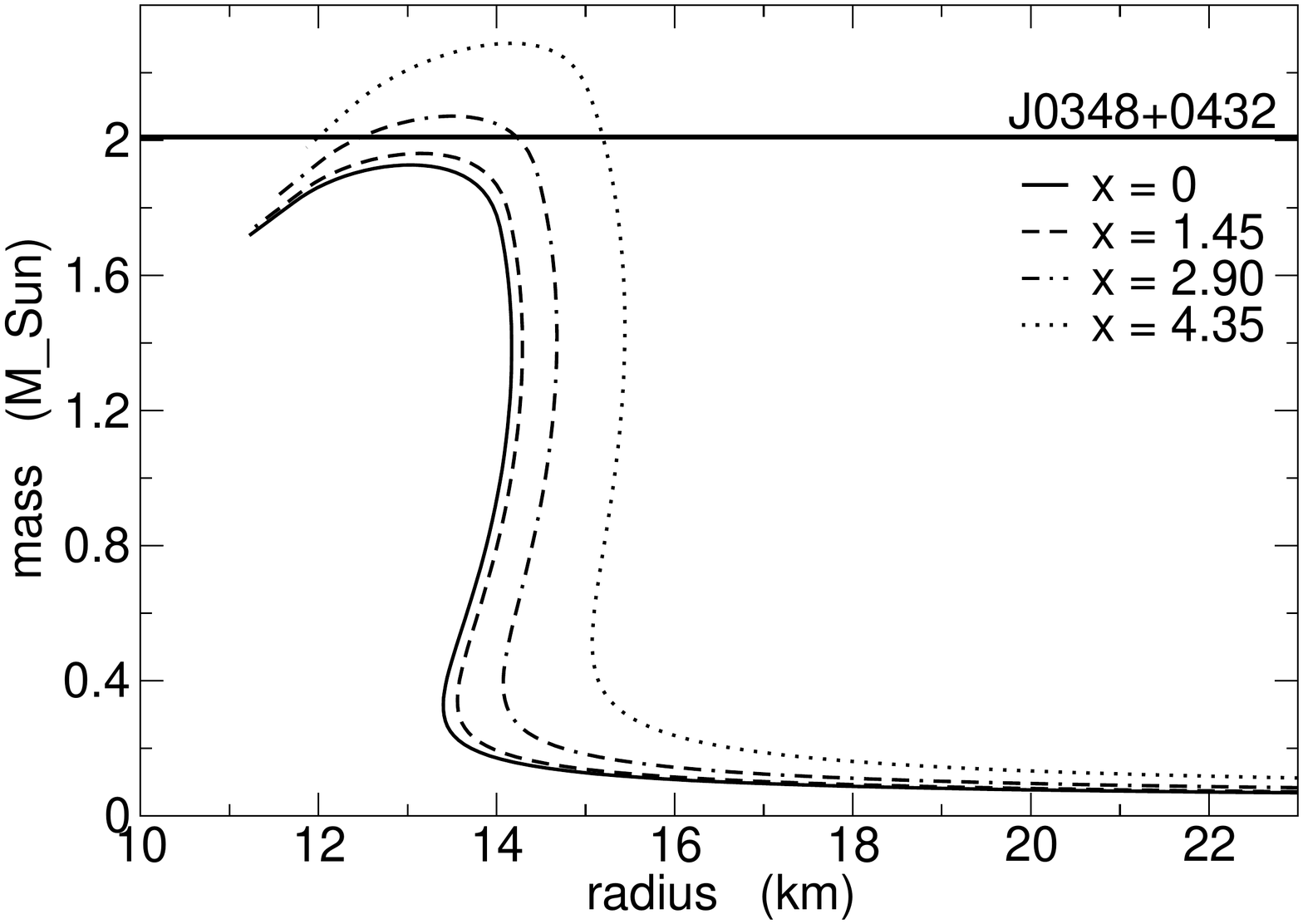}
\hskip 1.2cm
\includegraphics[height=8.4cm,width=8.4cm]{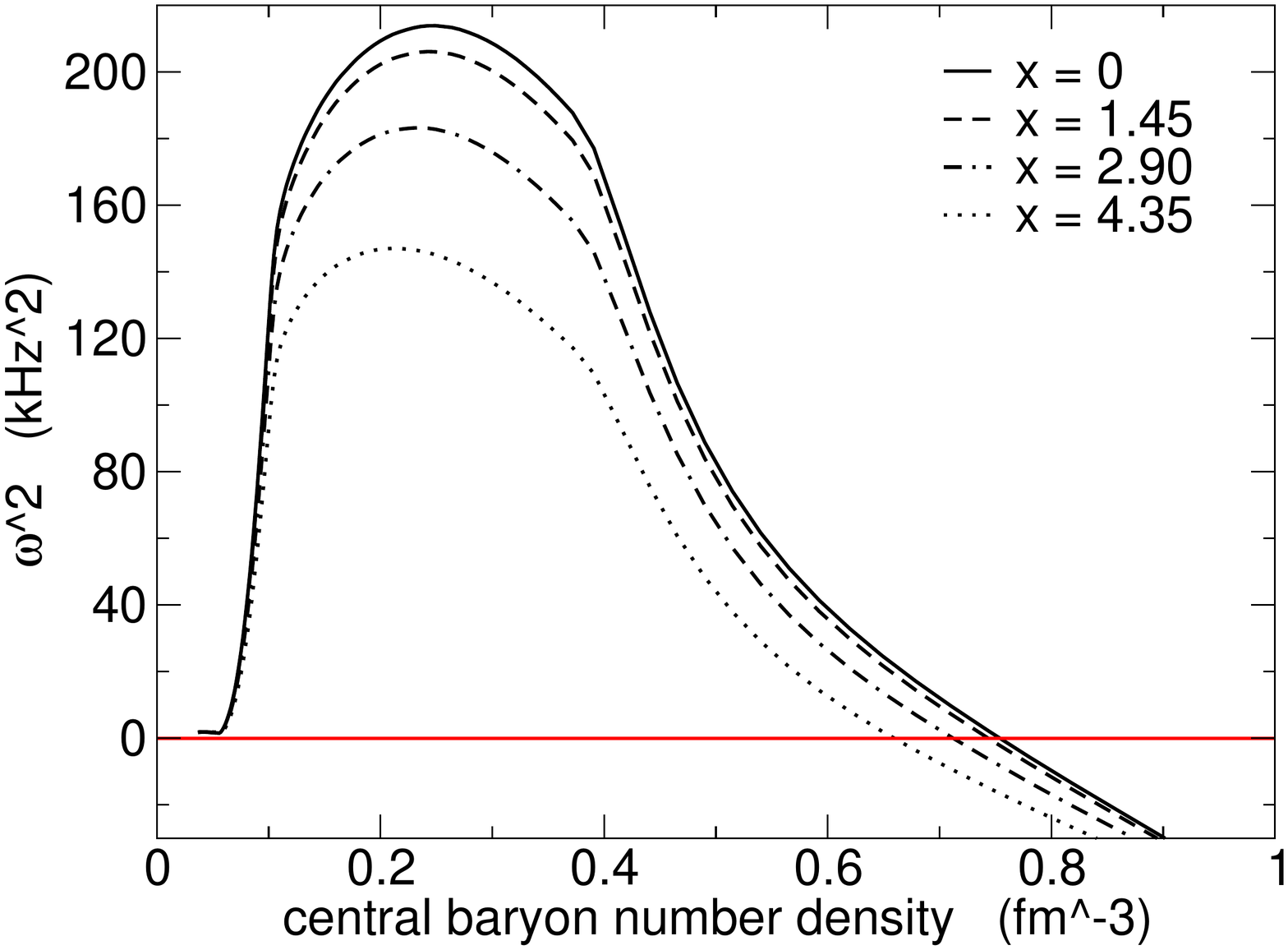}
\end{center}
\caption[]
{Mass-radius diagramm (upper pannel) and square of the oscillation frequency of the fundamental mode (lower pannel) for different values of $x$, defined in section 2. The parameters for the quark phase are: $m_s=100\;MeV$, $a_4=0.8$ and $B^{\frac{1}{4}}=200\;MeV$.}
\label{fig:m-r_1}
\end{figure}

\section{Results}
\noindent To calculate the frequencies in the linear approximation we first solve the equations of relativistic stellar structure. Then the discrete values of $\omega^2$ are found by matching the boundary conditions $\xi(r=0)=0$ and $\Delta P(r=R)=0$. The calculations were done with a 4th order Runge-Kutta algorithm with constant stepsize. A cubic spline interpolation scheme is used for the realistic EoS given in tabulated form. To test the validity of our program and eliminate the dependence on the interpolation part, a comparison was made with the results for the polytropic equation of state $p=\kappa\rho^{1+1/n}$ given in \cite{kokkotas}. Using the parameters $n=1$ and $\kappa = 100\,km$ we observe deviations in ${\omega_0}^2$ below 1\%, except for a single star with central energy denisty $\rho_c=5.65\,10^{15} g/cm^3$, for which the relative error in ${\omega_0}^2$ exceeds 13\%. In this case the star is close to the maximum mass configuration and the value of ${\omega_0}^2$ is almost zero, giving rise to the large relative error. We verified, that the code gives zero oscillation frequency at both extrema in the mass-radius diagram.

The integration of the eigenmode equation on a stellar background with the Maxwell phase transition can lead to positive values of $\omega^2$, even though $\frac{dM}{dn_c}<0$. In \cite{flores2} the question was risen, how to align this behavior with the known static stability criterion \cite{haenselbook}. While not considered in the present study, we argue that in the case of density jumps the eigenmode equation should not be applied through the discontinuity, but rather for each phase separately, with additional boundary conditions at the phase boundary.

\noindent In Figure ~\ref{fig:m-r_0} we show the mass-radius diagramm of neutral hybrid stars for different parameter choices of the quark matter EoS. We vary the value of the bag constant (upper pannel) and the $a_4$-parameter (lower pannel), which parameterizes corrections from the strong interaction. An increase in the bag constant yields higher critical densities for the phase transition. To comply with the mass limit the presence of the mixed phase is limited to a small region at the center. The associated frequencies are shown in Figure ~\ref{fig:osc_0}. At the onset of the phase transition there is a kink in the frequency curve. The critical central density, for which instability sets in, is shifted to lower values. The mass-radius and frequency diagramms for charged hybrid stars are given in Figure ~\ref{fig:m-r_1}. The Coulomb repulsion augments the masses and reduces the frequencies of the fundamental radial mode. This reduction is given at all values of central densities, but is more pronounced above $0.12 fm^{-3}$. The central densities associated with the maximum mass are shifted to lower values. In Fig. \ref{fig:m-r_1} we show results only for relatively small values of $x$, because at higher $x$ the central pressure approaches zero and the star structure should change. 

\section{Conclusion}
\noindent
We have constructed models of hadronic and hybrid stars consistent with recent observations \cite{antoniadis}. It was shown, that the presence of deconfined quarks at the center of hybrid stars reduces the oscillation frequencies of the fundamental radial mode. We have generalized the equation for radial eigenmodes to the case of charged stars. The general behavior is a decrease in the oscillation frequency at given central baryon density. The onset of instability is shifted to lower central densities. The departure from global charge-neutrality is accompanied by a substancial increase in the gravitational mass. This stems from the energy of the electric field and the Coulomb repulsion, which allows the star to contain more baryons. The mass increase allows to reconcile the observational constraint with soft equations of state, which are otherwise ruled out. Similar to the case of rotating equilibria, there exists a class of charged supermassive stars, for which there are no neutral stars with the same baryon number. 

\section{Appendix}
For completeness we give the oscillation equation \eqref{C59} in two different notations, encountered in the literature.
\subsection{Bardeen's form}
 In \cite{bardeen} \eqref{C59} was written in a form, which exhibits its Sturm-Liouville nature. With the renormalized displacement function $u=r^2e^{-\Phi_0}\xi$ \eqref{C59} reduces to
\begin{align}
&\frac{d}{dr}\left[\mathcal{P}\frac{du}{dr}\right]+\left[\mathcal{Q}+\omega^2\mathcal{W}\right]u=0\\
\mathcal{P}&=e^{\Lambda_0+3\Phi_0}r^{-2}\gamma P_0\\
\mathcal{Q}&=\left(\rho_0+P_0\right)r^{-2}{\Phi_0'}^2e^{\Lambda_0+3\Phi_0}-4r^{-3}{P_0}'e^{\Lambda_0+3\Phi_0}\notag\\&-8\pi\left(\rho_0+P_0\right)r^{-2}e^{3\Phi_0+3\Lambda_0}P_0\notag\\&-\left(\rho_0+P_0\right)r^{-6}e^{3\Phi_0+3\Lambda_0}{Q_0}^2\\
\mathcal{W}&=e^{3\Lambda_0+\Phi_0}r^{-2}\left(\rho_0+P_0\right)
\end{align}
Our $\mathcal{Q}$ agrees with the expression in \cite{rothman}, but is in contradiction to the expression in \cite{felice}, which we believe to be wrong. We find $\mathcal{Q}_{\,de\,Felice}=\mathcal{Q}+e^{\Lambda_0+3\Phi_0}Q_0{Q_0}''/\left(4\pi r^6\right)$.
\subsection{Gondek-Rosinska's form}
\noindent Chanmugam has written \eqref{C59} in \cite{chanmugam} as a system of two 1st order equations for $(\Delta P/P_0)'$ and $(\xi/r)'$. This is numerically advantageous in order to avoid the second derivative $\partial^2_{r\rho}P$. Similarly, Gondek-Rosinska et. al. have written \eqref{C59} in \cite{gondek} as a system for ${\Delta P}'$ and $(\xi/r)'$. This choice simplifies the outer boundary condition at the surface. We follow them and use the variable $\zeta=\xi/r$, which is equal to their $\xi$.

\begin{align}
\frac{d \zeta}{dr}&=-\frac{1}{r}\Big[3\zeta+\frac{\Delta P}{\gamma P_0}\Big]+\zeta{\Phi_0}'\notag\\
\frac{d\Delta P}{dr}&=\zeta\Big[\omega^2 e^{2\Lambda_0-2\Phi_0}\left(\rho_0+P_0\right)r-4{P_0}'+\left(\rho_0+P_0\right){\Phi_0'}^2 r\notag\\&-8\pi r\left(\rho_0+P_0\right)e^{2\Lambda_0}P_0-\left(\rho_0+P_0\right)r^{-3}e^{2\Lambda_0}{Q_0}^2\Big]\notag\\&-\Delta P\Big[{\Phi_0}'+4\pi r\left(\rho_0+P_0\right)e^{2\Lambda_0}\Big]
\end{align}

\acknowledgments
We thank J. Schaffner-Bielich, S. Schramm and R. P. Negreiros for fruitful discussions. We thank Irina Sagert for providing the EoS tables on her web page. This work was supported by the Hessian LOEWE initiative through the Helmholtz International Center for FAIR and the Helmholtz Graduate School for Hadron and Ion Research. I. M. acknoledges partial support from the grant NSH-215.2012.2 (Russia).

\end{document}